# Formation of early water oceans on rocky planets

Linda T. Elkins-Tanton[1]

[1]Massachusetts Institute of Technology, Department of Earth, Atmospheric, and Planetary Sciences, 77 Massachusetts Av., 54-824, Cambridge MA 02139, ltelkins@mit.edu

**Abstract**

Terrestrial planets, with silicate mantles and metallic cores, are likely to obtain water and carbon compounds during accretion. Here I examine the conditions that allow early formation of a surface water ocean (simultaneous with cooling to clement surface conditions), and the timeline of degassing the planetary interior into the atmosphere. The greatest fraction of a planet's initial volatile budget is degassed into the atmosphere during the end of magma ocean solidification, leaving only a small fraction of the original volatiles to be released into the atmosphere through later volcanism. Rocky planets that accrete with water in their bulk mantle have two mechanisms for producing an early water ocean: First, if they accrete with at least 1 to 3 mass% of water in their bulk composition, liquid water may be extruded onto the planetary surface at the end of magma ocean solidification. Second, at initial water contents as low as 0.01 mass% or lower, during solidification a massive supercritical fluid and steam atmosphere is produced that collapses into a water ocean upon cooling. The low water contents required for this process indicate that rocky super-Earth exoplanets may be expected to commonly produce water oceans within tens to hundreds of millions of years of their last major accretionary impact, through collapse of their atmosphere.

# 1 Introduction

Water and carbon are widely thought to be necessary for the development and maintenance of life. The Earth is the only terrestrial planet in our solar system with liquid water oceans, though they may have existed in the past on both Venus and Mars. Volatile components such as water and carbon dioxide, despite their relative scarcity in terrestrial planets, have a disproportionate influence on the processes and outcomes of planetary differentiation and on habitability. Here I present models for planetary formation that span a range of initial compositions indicated by the range of meteorite samples found on Earth, and investigate the processes that might give rise to water oceans on planets of 1 to 5 Earth masses.

The terrestrial planets in our solar system have differentiated to include a silicate mantle, metallic core, and in some cases, a gaseous atmosphere. Differentiation of metal from silicate is proposed to require melting of the metal and likely of the silicate portion of the planet as well. Heat for this melting comes from the kinetic energy of accretion; giant accretionary impacts are expected to create partial- or whole-mantle magma oceans (*e.g.*, Wood et al., 1970; Canup, 2008). The molten state is a good initial condition for forward modeling, since the chemistry and physics of solidification are relatively tractable.

Evidence from ancient zircon minerals indicates that the Earth attained clement surface conditions within about 150 million years of the first solids condensing in the solar system (Valley et al., 2002), perhaps less than 100 million years after the Moon-forming impact (Touboul et al., 2007). Here I show that a water ocean may be obtained within that time period on the Earth (and similarly for super-Earths) through two primary mechanisms: First, enrichment of water in a solidifying magma ocean to the point that excess water is extruded onto the Earth's surface at the end of silicate mineral solidification. Second, a thick supercritical fluid and steam atmosphere may form a water ocean when it cools sufficiently to stabilize water (Abe & Matsui, 1985).

# 2 Planetary compositions

Meteorites provide a range of compositions from which terrestrial planets may have been built. Meteorites that came from primitive or largely undifferentiated parent bodies can contain up to 20 mass% water, while those from differentiated parent bodies contain less than 3 mass% water (Jarosewich, 1990) (Figure 1). As a comparison, the total of liquid and frozen water on Earth's surface is about $1.4 \times 10^{21}$ kg, while the mass of the Earth's mantle is $6 \times 10^{24}$ kg. In bulk, then, the silicate Earth plus surface water composition 0.02 mass% water, or 200 ppm water. There is an additional undetermined quantity of water in the Earth's silicate mantle, as well, so 200 ppm is an unrealistic minimum bulk water value for the sum of the silicate mantle and hydrosphere of the Earth.

Terrestrial planets may obtain the bulk of their water and carbon during the giant impacts of accretion; this is the hypothesis I am exploring here. Planets may also have accreted drier and obtained volatiles from later, surficial additions during the tailing off of accretion and clearing out the inner solar system (*e.g.*, Ringwood, 1981; Wänke, 1981). I begin these models, therefore, with 100 ppm (0.01 mass%) of water in the bulk silicate planet, and the highest initial water content considered is 3 mass%.

For the remainder of the silicate portion of the planet the models use an estimate of the bulk silicate Earth from Hart & Zindler (1986). Though planets around other stars are likely to have different bulk compositions (and indeed even Mars appears to have more iron oxide in its mantle than does Earth [Bertka & Fei, 1997]) the processes of solidification and volatile enrichment are likely to be similar.

## 3 Magma ocean solidification

The pressure gradient in the magma ocean, determined by the mass of the planet, controls the processes of solidification. The adiabatic gradient in a convecting magma ocean is steeper than the solidus of the magma, and so as the adiabat moves to lower temperatures through secular cooling, it intersects the solidus first at depth (Solomatov, 2000; Elkins-Tanton, 2008, and references therein). Almost all minerals that solidify from a bulk Earth magma composition are also denser than their coexisting magma. Solidification will therefore proceed from the bottom upward, with increasingly evolved magmas lying above growing layers of solid silicate minerals.

Because water and carbon are only minimally incorporated in these minerals, the magma is increasingly enriched in water and carbon compounds as solidification proceeds. These models calculate the compositions of solidifying minerals, beginning at depth with the post-perovskite phase (Figure 2), and record the composition of the evolving magma ocean liquids. For details of these models, see Elkins-Tanton (2008).

The first mechanism for creating a water ocean is enrichment of magma ocean liquids with water while depleting them of other oxides through mineral solidification; water enrichment thus depends upon the composition of the solidifying minerals. The minerals shown in Figure 2 are nominally anhydrous, that is, water is not an integral part of their crystal structure. As solidification proceeds toward the planetary surface, pressure declines, and water builds up in the magmas, water-bearing minerals (not included in Figure 2) will become stable. Water is either incorporated into solidifying hydrous minerals, or remains free to form an ocean.

The minerals chlorite, serpentine, and other hydrated silicates contain not more than 10 to 15 mass% water (Deer et al., 1996). Other, rarer non-silicate minerals such as epsomite and borax can contain as much as 50 mass% water (Deer et al., 1996). The bulk composition of late-stage magma ocean liquids will determine which minerals are stable. Though these last few percent of mineral precipitation cannot be modeled accurately without experimental work, late-stage magma ocean liquids will eventually become so enriched with water as to be effectively super-critical water oceans. Silicate solidification will end because the few oxides remaining are soluble in what is then effectively a hot water ocean.

Based upon the maximum water content of hydrous minerals that may solidify from the last dregs of the magma ocean, in the absence of experimental data to constrain models, magma oceans that attain more than 15 to 25 mass% water in their final percent of liquid may be expected to produce such a water ocean. For planets between 1 and 5 Earth masses, this requires they begin with at least 1 to 3 mass% water in their bulk mantle composition (Figure 3). The massive atmospheres produced by such initial water contents may indicate that they are unreasonably high; see Section 4, Atmospheric Growth. Further, 3 mass% is the maximum water content found in meteorites from differentiated parent bodies, and is therefore a maximum likely initial water content for any terrestrial planet in our solar system.

Some evidence exists that, at pressures near the bottom of Earth's mantle, magma becomes more dense than coexisting minerals and solidification would proceed downward from a septum (Mosenfelder et al., 2007). If deep magma oceans solidify from the top down, their volatiles would be trapped in the planetary interiors rather than released onto the surface. In the calculations for this paper I assume all magma oceans solidify from the bottom upwards. If a planet had a whole-mantle magma ocean and solidified upward only from a mid-mantle septum, the values for oceanic depth presented here would be too great.

Planetary mantles are likely, however, to be processed through a series of magma oceans as accretionary impacts build the planet. The larger the planet becomes, the less likely a given impact can melt the entire depth of the mantle (even the putative Moon-forming impact with the Earth may only have melted it to 2,000 km depth [Canup, 2008]). The processes of degassing and enrichment of magma ocean liquids toward the planetary surface, therefore, will accumulate water on the planetary surface in the same way a larger magma ocean that solidifies upward would do, but with the losses produced by the impact of each new accreting planetesimal. These impacts might be expected to blow off about half of the surface volatiles (Genda & Abe, 2003).

**4 Atmospheric Growth**

Water in excess of the saturation limit of the evolving magma ocean is assumed in these models to be degassed into a growing atmosphere (Elkins-Tanton, 2008). New results indicate that a planetary atmosphere will not likely grow gradually, but will degas catastrophically toward the end of solidification (Suckale & Elkins-Tanton, 2010), but these models do not incorporate catastrophic degassing. Degassing early in solidification is minimal in these models, however, and the catastrophic model will not significantly alter the results shown here.

Between 60 and 99 mass% of the original water in the planetary magma ocean is degassed into the growing atmosphere because so little water can be incorporated into the nominally anhydrous minerals of the planetary mantle. Even very small initial water contents, therefore, can produce substantial steam atmospheres (see the atmospheric pressures given in Figure 3). While a super-critical water ocean produced on the planetary surface directly from magma ocean liquids as described in the section above requires more than ~1 mass% water in the bulk magma ocean, a whole-mantle magma ocean with 0.1 mass% (1,000 ppm) water initial produces a steam atmosphere of hundreds of bars on planets from 0.5 to 5 Earth masses. The steam atmosphere is the major volatile reservoir for the planetary surface (Figure 4).

Elkins-Tanton (2008) showed that, for Earth, surface temperatures near the super-critical point of water are reached within tens of millions of years of the initiation of a magma ocean. For super-Earths, the timescale may be a factor of ten longer. When the surface temperature cools below the critical point (approximately 647K and 220 bars) the supercritical fluid and steam atmosphere will collapse into a surface ocean.

Planets from 0.5 to 5 Earth masses that have a bulk composition including 0.01 mass% water, processed through one or more magma oceans, produce a surface ocean of hundreds of meters depth upon cooling and condensation of a steam atmosphere. If these planets began with 0.1 mass% water, the global ocean resulting from atmospheric collapse would be 1 to 10 kilometers deep, and beginning with 0.5 mass% water results in global water oceans tens of kilometers deep (Figure 5).

## 5 Conclusions

Although producing a water ocean directly from progressive solidification of a magma ocean and attendant enrichment of water is possible, it requires concentrations of water that are considered unlikely for Earth's formation conditions. These levels of water (above 1 mass%, likely near or above 3 mass% for larger super-Earths) are possible; however, the more probable source for early water oceans is the collapse of the planet's steam atmosphere.

Within tens of millions of years of the initiation of a magma ocean the planetary mantle will be solid and compositionally stable, and the planet's surface temperature will have passed the critical point of water and cooled into the liquid water stability field (Abe, 1997; Elkins-Tanton, 2008). Planets that began with as little as 0.1 mass% or less of water are likely to form a water ocean of hundreds of meters depth. These models do not include atmospheric escape, and so the longevity of such oceans remains a question.

Though these oceans may not persist over billions of years on smaller planets against the processes of atmospheric escape and continuing impact blow-off, their early existence may be important for the development of life; larger planets would be expected to retain their oceans more effectively. The low initial water content required indicates that rocky super-Earth exoplanets commonly may be expected to produce both a liquid water ocean directly from a solidifying magma ocean, and a far larger ocean from later atmospheric collapse.

**Acknowledgements**

This work was improved by reviews and was funded by an NSF CAREER grant to the author through the Astronomy Program.

**Figure Captions**

**Figure 1.** Both primitive (chondritic) and differentiated meteorite compositions can contain significant water and carbon. The range of water used in this paper (inset) includes meteorites that are likely building blocks for terrestrial planets. Data from Jarosewich (1990).

**Figure 2.** Simplified mineral assemblages expected in the silicate mantles of planets with near-Earth composition. These interior minerals can contain limited fractions of OH, as shown, and accept these hydoxyl molecules in proportion to the water content of the coexisting magma, from which they are crystallizing. Water beyond the capacities of these minerals will remain in the magma.

**Figure 3.** Pressure of steam atmosphere produced by degassing magma ocean as a function of planet mass and initial water content. Shaded region indicates the range of water contents of achondritic (differentiated) meteorites, the likely building blocks of terrestrial planets. Dashed lines indicate the initial bulk water content in a whole-mantle magma ocean. Bold lines indicate the water content of the final 1 vol% of magma ocean liquids when solidification of the planet is almost complete. For example, a planet with 3 Earth masses that begins with 3 mass% water in its whole-mantle magma ocean will end with the final 1vol% of magma ocean liquids containing between 15 and 25 mass% water, possibly sufficient to form a water ocean almost immediately. Current locations of terrestrial planets are marked.

**Figure 4.** Schematic timeline of planetary degassing into oceans and atmospheres. The vertical axis represents the fraction of volatiles present in the magma ocean before solidification begins that are degassed (1 = all initial volatiles degassed; none remaining in planetary interior). The bold solid line represents larger planets and lower initial volatile contents in magma oceans (0.01 mass% or less), while the bold dashed line represents smaller planets and higher initial volatile magma oceans (see Table 3 in Elkins-Tanton, 2008). The time to magma ocean solidification depends primarily upon initial volatile content, which along with planetary mass controls the time required for solid-state mantle overturn - a process driven by density gradients within the solid mantle following the magma ocean stage (Solomatov, 2000; Elkins-Tanton, 2008} - and the length of mantle convection suppression that follows overturn to stability. By far the greatest fraction of volatiles are released onto the planetary surface during solidification.

**Figure 5.** Log of resulting water ocean depth in meters produced by a collapsed degassed atmosphere of a planet of given mass and initial bulk mantle water content. Water contents far less than 1 mass% would likely produce near-planetary water oceans within tens of millions of years of the last accretionary impact.

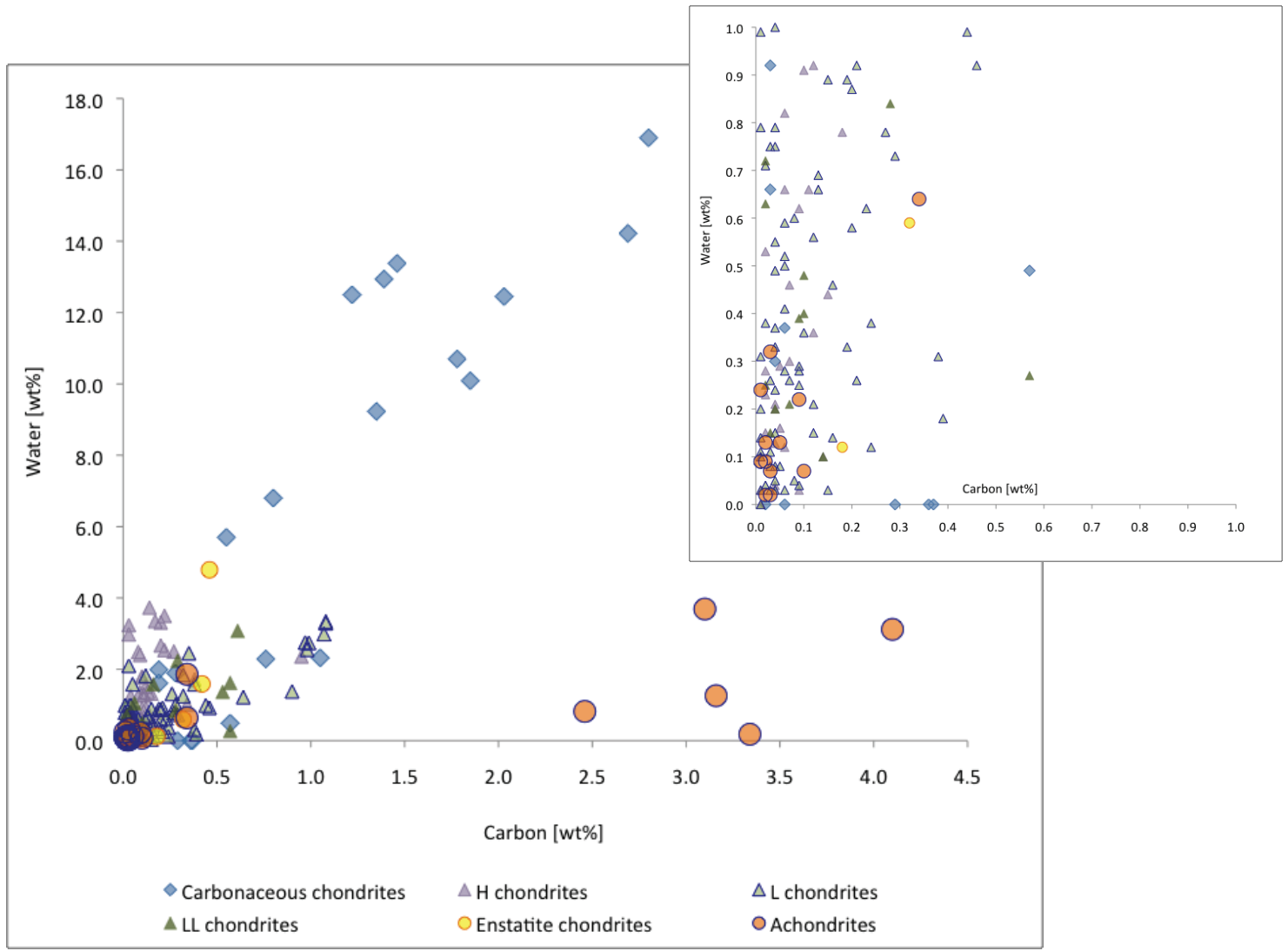

Elkins-Tanton Figure 1

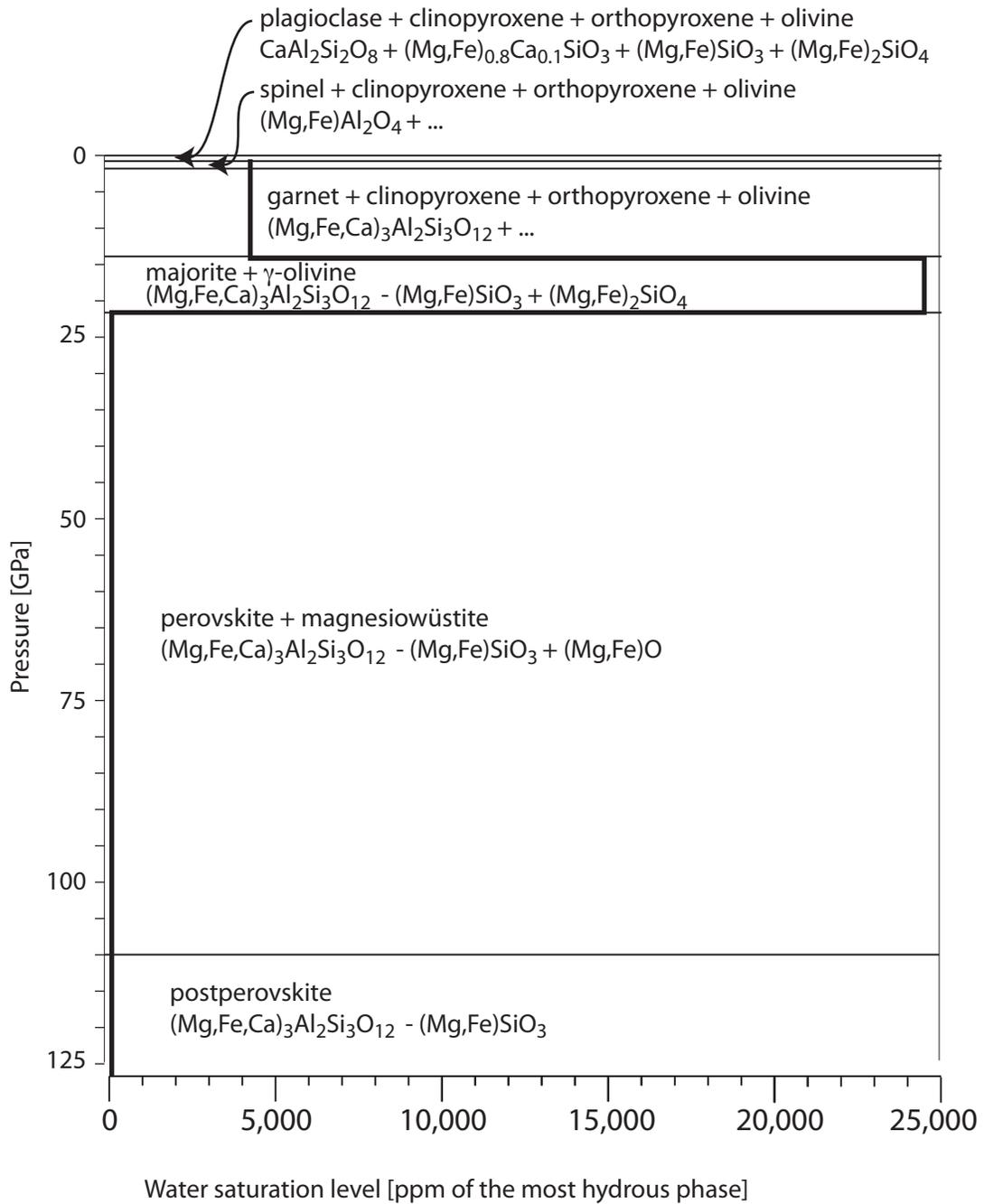

Elkins-Tanton Figure 2

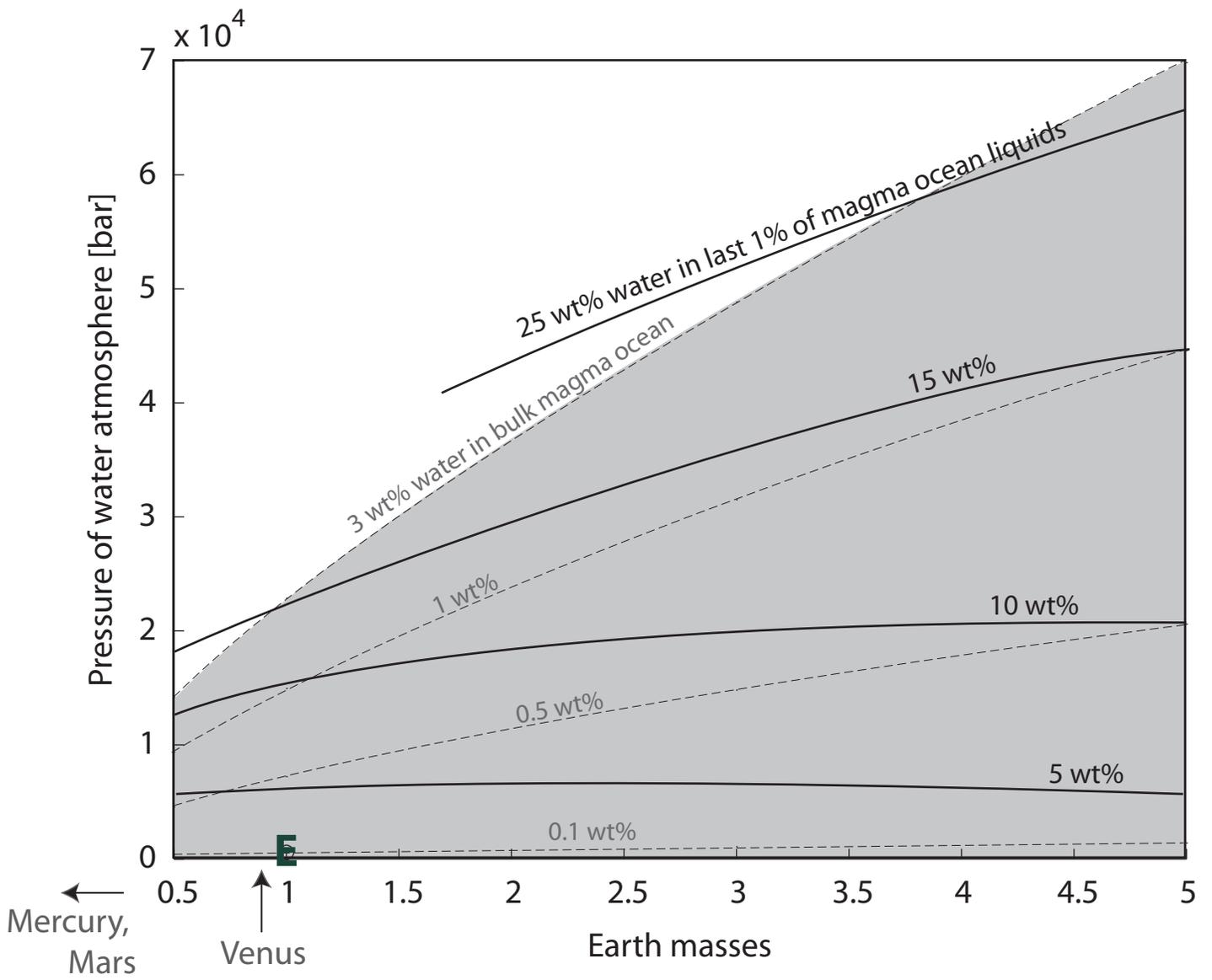

Elkins-Tanton Figure 3

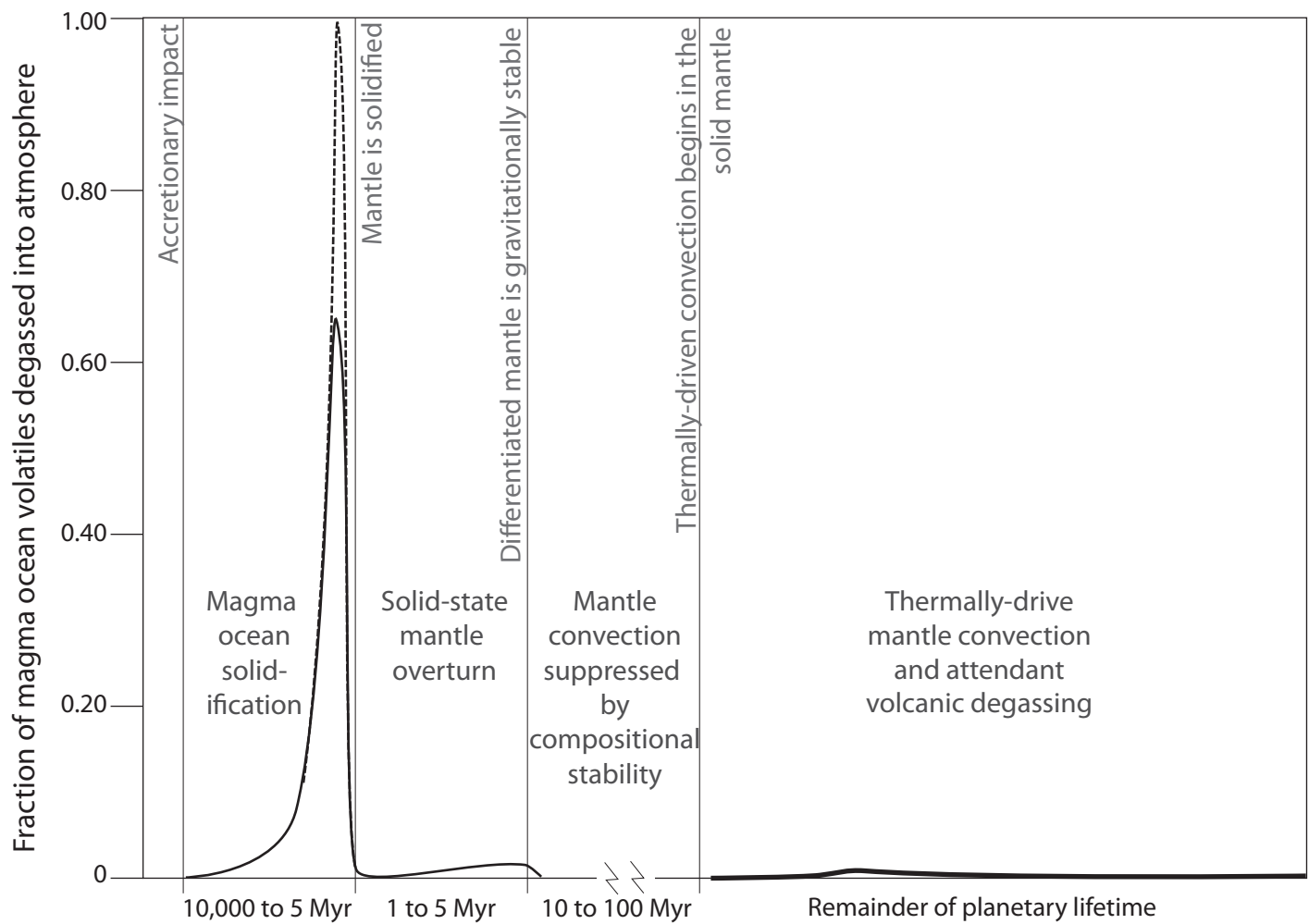

Elkins-Tanton Figure 4

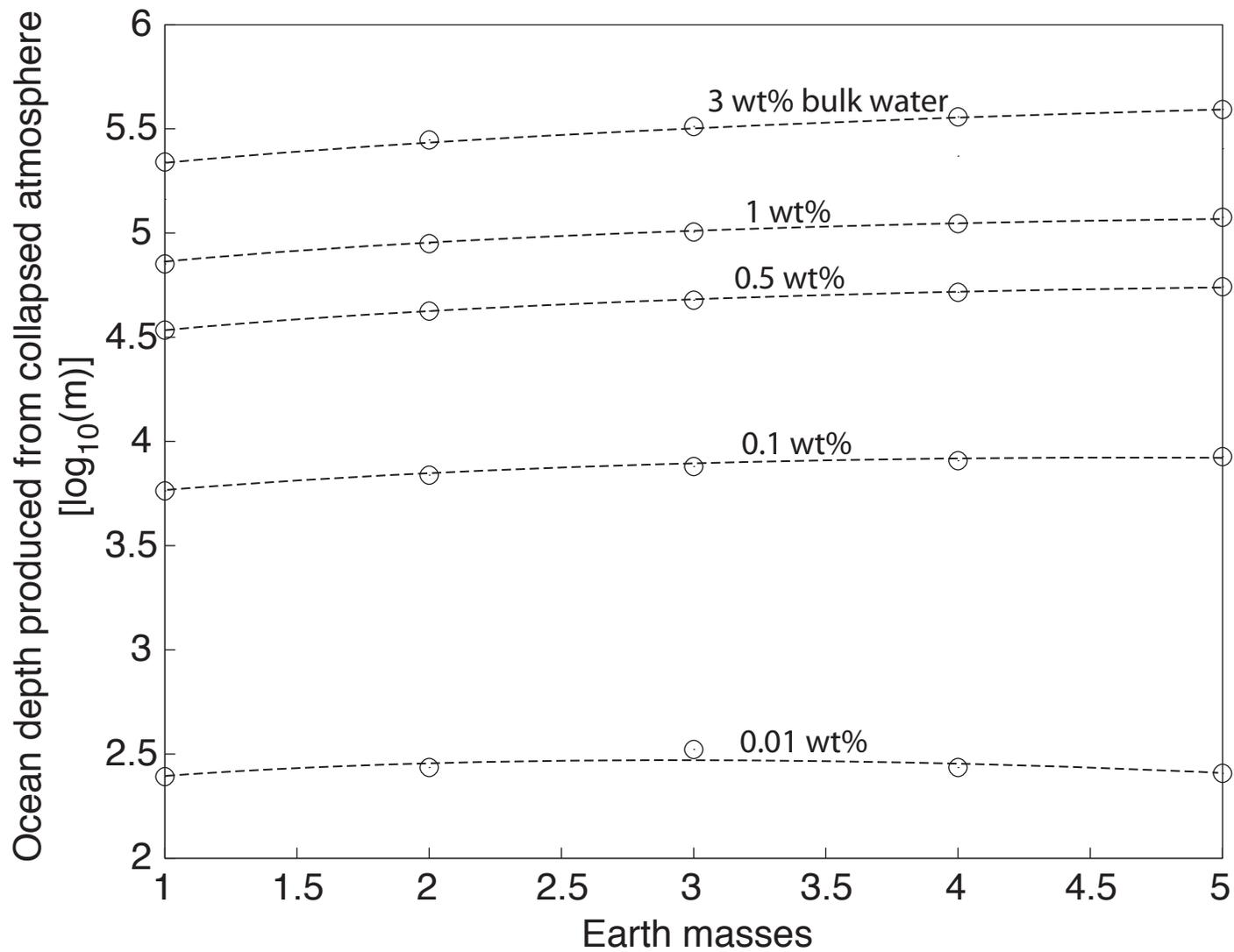

Elkins-Tanton Figure 5